\documentclass[preprint,showpacs,preprintnumbers,amsmath,amssymb,aps]{revtex4}

\usepackage{graphicx}
\usepackage{dcolumn}
\usepackage{bm}
\begin{document}

\title{Glassy features of a Bose glass}

%
\author{\firstname{Pierfrancesco} \surname{Buonsante}$^1$}
%
\author{\firstname{Francesco} \surname{Massel}$^1$}
%
\author{\firstname{Vittorio} \surname{Penna}$^1$}
%
\author{\firstname{Alessandro} \surname{Vezzani}$^2$}
\affiliation{$^1$Dipartimento di Fisica and C.N.I.S.M., Politecnico di Torino Corso Duca degli Abruzzi 24, I-10129 Torino (ITALIA)}
\affiliation{$^2$Dipartimento di Fisica, Universit\`a degli Studi di Parma and C.N.R.-I.N.F.M., Parco Area delle Scienze 7/a, I-43100 Parma (ITALIA)}
%
%
%

\begin{abstract}
We study a two-dimensional Bose-Hubbard model at zero temperature 
with random local potentials in the presence of either uniform or binary disorder. Many low-energy metastable configurations are found 
with virtually the same energy as the ground state's.
These are characterized by the same blotchy pattern of the 
---in principle complex--- non-zero local order parameter as 
the ground state. Yet, unlike the ground state,
each island exhibits an overall random independent phase. 
The different phases in different coherent islands could provide a further
explanation for the lack of coherence observed in experiments on Bose glasses.
\end{abstract}
\pacs{03.75.Lm, 05.30.Jp, 64.60.Cn}

\maketitle

\section{Introduction}

The problem of interacting bosons hopping across the sites of a disordered lattice has been addressed since the seminal paper on the Bose-Hubbard (BH) model by Fisher and co-workers \cite{A:Fisher}. There it was observed that the presence of disorder (in the form of random on-site potentials) enriches the phase diagram of the homogeneous model adding one further {\it Bose-glass} (BG) phase to the  {\it superfluid} (SF) and {\it Mott-insulator} (MI) phases arising from the competition between on-site repulsion and kinetic energy.
%
Several studies followed Ref.~\cite{A:Fisher}, employing various techniques and possibly addressing different realizations of disorder. 
The latter include models with diagonal disorder in one-dimensional
(1D) lattices \cite{A:Scalettar91,A:Freericks96,A:Rapsch99,A:Gimper05,A:DeMart05,
A:Yuka07,A:Mering07}, and 
in 2D lattices \cite{A:Krauth92}, \cite{A:Singh94}, 
as well as
off-diagonal disorder in 1D lattices
\cite{A:Sengupta99}, \cite{A:Balab05}
and in 2D lattices \cite{A:Bernad02}, \cite{A:Prokov04}.

Recently, the interest in this subject has been boosted by the impressive advances in the field of cold atom trapping. Indeed, the BH model can be realized in terms of ultracold bosonic atoms trapped in optical lattices
where disorder can be engineered with different experimental techniques.
%
A speckle field can be superimposed otherwise an otherwise regular lattice, resulting in random modulations of local lattice parameters 
\cite{A:Lye}, \cite{A:Schulte}. A similar (pseudo)randomness can be obtained by superimposing two optical lattices with uncommensurate lattice constants \cite{A:Damski2003}, \cite{A:Fall07}. Also, a regular optical lattice can be loaded with a small amount of a second (possibly fermionic) species with reduced mobility resulting from a quench in their hopping amplitude. Due to their interaction with the bosonic species, the  atoms of the second species act as randomly localized scatterers \cite{A:Gavish,A:Ospelkaus}.

It has been shown that a mean-field Gutzwiller approach captures
the phases of the Bose-Hubbard model also in the presence of disorder 
\cite{A:Sheshadri1995,A:disMF1d,A:RandhopMF,A:disMF2d}. Here we show that such an approach also highlights some features of the BG which support the reference to glasses in the name of such a phase. Indeed, the phase diagram of the disordered BH model is the subject of many investigations where the BG is usually characterized in terms of {\it gaplessness}, compressibility, superfluid and condensate fraction.
Conversely, a few works focus on the
typical features characterizing a glassy system \cite{A:Krutitsky}, 
such as, in particular, the
exixtence of a complex energy landscape with a large number of local
minima which should play a fundamental role in the relaxation dynamics.
We expose the existence of many low-lying {\it metastable} states which differ from the BG ground-state essentially only for the phase pattern of the order parameter. More specifically the BG phase is characterized by patches of finite local order parameter separated by regions where the local order parameter vanishes. This speckled pattern of the local order parameter results in a finite condensate fraction and a vanishing superfluidity \cite{A:disMF1d}. As we are going to illustrate, many states exist having the same distribution of  local densities and (absolute value of)  order parameters. The lowest energy is attained by the configuration where the phases of all the local order parameters are aligned along the same (arbitrary) direction. However, it turns out that many extremely stable configurations exist where the phase is aligned within each order-parameter patch, but different patches have different phases. The energy of such {\it metastable} configurations is only sligthly larger than that of the ground state. This scenario is clearly reminiscent of glassy systems.
\section{The system}
The BH model is currently realized in laboratories in terms of
ultracold bosonic gases trapped in optical lattices
\cite{A:Lye,A:Schulte,A:Damski2003,A:Fall07,A:Gavish,A:Ospelkaus}, \cite{A:Clement05}.
Its Hamiltonian reads
\begin{eqnarray}
\label{E:BHH}
H=\sum_{i=1}^M \left[U_i \frac{n_i-1}{2} + v_i \right]n_i
- \! \sum_{\langle i,h \rangle} J_{i, h} \left(a_i^\dag a_h + h.c. \right)
%
\end{eqnarray}
where $M$ is the lattice size, $a_i$, $a_i^\dag$ and $n_i$ destroys, creates and counts bosons at lattice site $i$, respectively. The local boson-boson interaction $U_i$, on-site potential $v_i$ and hopping amplitude across neighbouring sites $J_{i,h}$ can be tuned by varying controllable experimental parameters such as the atomic scattering length (via Feshbach resonance) and the strength and setup of the electromagnetic fields giving rise to the optical lattice. 
Disorder can be in principle introduced in all these parameters.
While the case with random $v_i$ has been mostly studied,
several works have addressed models with random $U_i$ or $J_{i,h}$ and the
phase diagrams thereof \cite{A:Gimper05,A:Prokov04,A:Balab05,A:RandhopMF}.
For the sake of simplicity, in the following we will assume the local (repulsive) interactions and the hopping amplitudes to be constant throughout the lattice, $U_i=U=1$, $J_{i,h}=J>0$ (that is, the on-site repulsive strength is our energy scale). Also, being interested in the phases of the system, we will consider a situation as closer to the thermodynamic limit as possible. Hence we will assume periodic boundary conditions, and that the harmonic confining potential typical of experimental system is so weak to be safely ignored.
In summary, the only site dependent quantity in our system will be the local potential $v_i$. We will consider two realization of disorder: local potentials uniformly distributed in $[-\Delta, \Delta]$ and {\it binary} disorder. In the latter case $v_i$ assumes the values $\Delta$  and $0$ with probabilities $p$ and $1-p$.

\subsection{Zero-temperature phase diagram}
Let us now briefly recall the main features of the phase diagram of Hamiltonian (\ref{E:BHH}). The different phases are characterized by the properties of the ground-state $|\Psi\rangle$ or of the low-lying sector of the spectrum. Standard quantities used to characterize the phase diagram are the energy gap between the ground and the first-excited state, the compressibility 
$\kappa = M^{-1}\partial_{\mu} {\cal N}$, where $\cal N$ is the total boson population controlled by chemical potential $\mu$, the condensate fraction
$f_{\rm C}$, i. e. the  maximal eigenvalue \cite{A:Roth}
of one-body density matrix 
$\rho_{i,h}= \langle \Phi |a_i^\dag a_h|\Phi\rangle$,
the SF fraction $f_{\rm S}$ measuring the response of the system
under the action of an infinitesimal
to a phase twist or boost (see, e. g., \cite{A:disMF2d}).

The phase diagram is usually drawn in the $\mu/U - J/U$ plane, and it 
is mostly taken by the SF phase, characterized by a gapless spectrum,
finite compressibility $\kappa >0$, finite condensate and SF
fractions $f_{\rm S}, \, f_{\rm C} >0$. The small-$J/U$ region of the 
$\mu/U - J/U$ plane is occupied by a series of MI lobes, where the system is characterized by a gapped spectrum and vanishing $\kappa$, $f_{\rm S}$ and 
$f_{\rm C}$. In the presence of disorder further phases may appear
\cite{A:Sengupta99}, the best-known being the BG. This was originally characterized as an insulating (non-SF) yet compressible (gapless) phase
\cite{A:Fisher,A:Scalettar91}. Later it was shown that, at least 
at the mean-field level, the BG is marked
%
%
by finite $\kappa$ and $f_{\rm C}$ and 
vanishing $f_{\rm S}$ \cite{A:disMF1d}.

\subsection{Gutzwiller Mean-Field Approximation}
The Hilbert space of Hamiltonian (\ref{E:BHH}) is infinite even on a finite system. However, one can take advantage of the total number conservation arising from the commutation of $H$ and the total number operator $N=\sum_i n_i$. This allows to address the properties of the BH Hamiltonian separately in each fixed-number Hilbert subspace, whose size is $d({\cal N},M)=\binom{{\cal N}+M-1}{{\cal N}}$ for ${\cal N}$ bosons on a size-$M$ lattice. Such a size makes exact diagonalization prohibitive also for relatively small lattices at fillings of the order of unity. Larger systems can be addressed by resorting to Complex and computationally demanding numeric simulations, such as quantum Monte-Carlo, Density-Matrix Renormalization Group and Time-Evolving Block Decimation algorithms. 

An alternative approach giving satisfactory qualitative results at much lower computational cost is provided by the {\it site-decoupling} (Gutzwiller) mean-field approximation. Basically, it relies on the posit $a_j^\dag a_k = a_j^\dag \alpha_k + \alpha_j^* a_k - \alpha_j^* \alpha_k$, where $\alpha_j=\langle a_j\rangle$ and $\langle \cdot \rangle$ denotes expectation value on the ground-state $|\Psi\rangle$ of the system. As a result, the mean-field BH Hamiltonian becomes the sum of ``decoupled'' on-site terms, ${\cal H}=\sum_j {\cal H}_j$, 
\begin{eqnarray}
{\cal H}_j= \frac{U}{2}n_j (n_j-1) + \left(v_j-\mu\right) n_j
-J \left(a_j^\dag \gamma_j+a_j \gamma_j^*\right)
%
%
\label{E:BHmf}
\end{eqnarray}
where $\gamma_j = \sum_j A_{j\,h} \alpha_h$. This in turn results into the mean-field ground-state being a product of on-site states,
\begin{equation}
|\Psi\rangle \approx \prod_{j=1}^M |\psi_j\rangle, \quad 
|\psi_j\rangle = \sum_{n=0}^\infty c_{j\,n} \left(a_j^\dag\right)^n |\Omega\rangle,
\label{E:GSmf}
\end{equation}
where $|\Omega\rangle$ is the vacuum state, $a_j |\Omega\rangle =0$.
Note that the constraint on the $\gamma_j$'s preserves some degree of coupling among neighbouring sites. Also, we recall that finding the ground-state of Hamiltonian $\cal H$ is equivalent to finding the ground-state (lowest-energy fixed-point) of a set of coupled nonlinear dynamic equations more general than the discrete Gross-Pitaevskii Equations 
\cite{A:Jaksch2,A:Snoek07}.
Such ground-state is fully determined by the set of {\it on-site order parameters} $\{\alpha_j\}$ such that 
$\langle \psi_j|a_j|\psi_j\rangle = \alpha_j$, where $\psi_j$ is the ground-state of the on-site mean-field Hamiltonian (\ref{E:BHmf}). 
Quantities $f_{\rm C}$ and $f_{\rm S}$, as well as the other quantities generally employed in the characterization of the different phases of
the model, are straightforwardly determined from once set $\{\alpha_j\}$
is known (see, e.g., \cite{A:disMF2d}).

We recall that, unlike Eq. (\ref{E:BHH}), the mean-field Hamiltonian does not commute with the total number of bosons: $[{\cal H},N]\neq 0$. Hence the chemical potential $\mu$ appearing in the local mean-field Hamiltonians (\ref{E:BHmf}) has to be properly set in order to attain the desired boson population ${\cal N}=\sum_j \langle\psi_j | n_j|\psi_j\rangle$.

\section{Results}
In the following we address a BH model on a $50\times50$ lattice
with periodic boundary conditions and $U_j=U>0$, $J_{j\,h} = J>0$.
%
As to the on-site
potentials $v_j$, we consider two different realizations of disorder, 
namely 
random potentials uniformly distributed in the interval $[-\Delta,\Delta]$,
with $\Delta=0.2$ and potentials assuming the value $\Delta=0.5$ with probability $p\approx 5\cdot10^{-2}$.

\begin{figure}[t!]
\begin{center}
\includegraphics[width= 18cm]{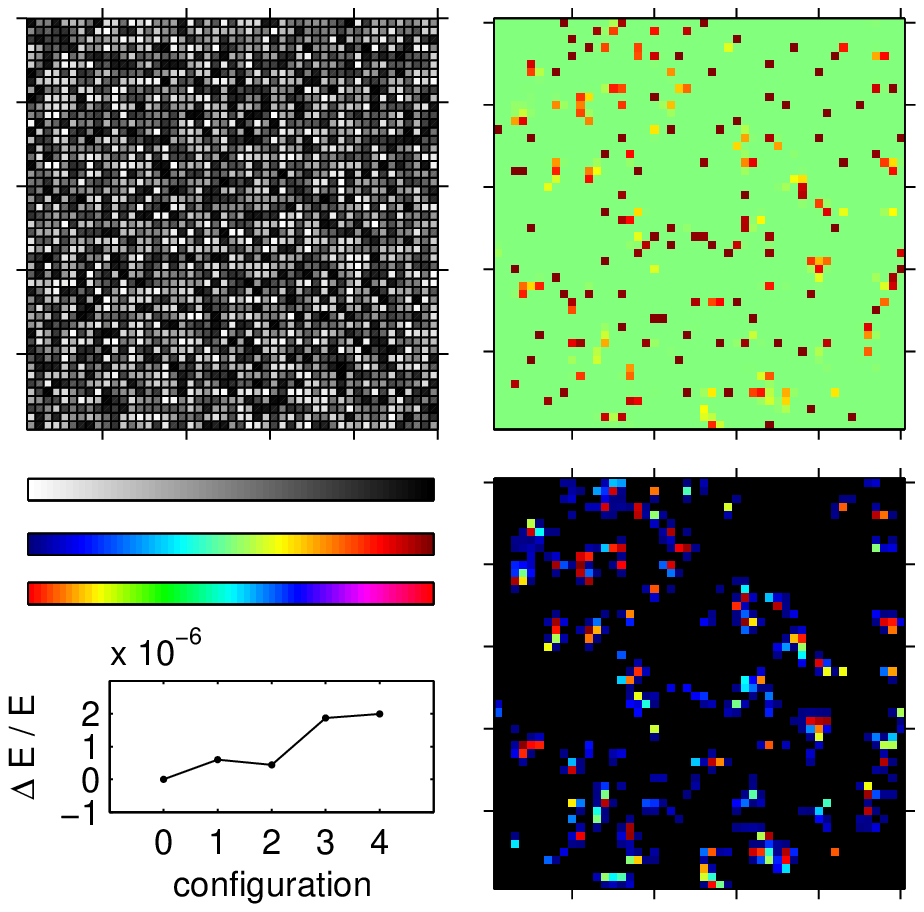}
\caption{\label{F:fig1}
Uniformly distributed random potential ($\Delta/U = 0.2$) on a $50 \times 50$ lattice with periodic boundary conditions containing $N=2650$ bosons with hopping amplitude $J/U=6\cdot 10^{-3}$. Upper left panel: random local potentials; Upper right panel: $\langle n_j\rangle$; Lower right panel: $|\langle a_j \rangle|^2$; The colormaps for these data are shown in the lower left quadrant of the figure. From top to bottom: local potentials, local density and order parameters (color ranges $[-0.2 U, 0.2 U]$ and $[0, 0.527]$, respectively), local phases, referring to Fig. \ref{F:fig2} (color range $[0,2 \pi]$). The plot below the colorbars shows the relative energy differences between the real (0) and complex  configurations (from 1 to 4, displayed in fig. \ref{F:fig2}). }
\end{center}
\end{figure}

For both the realizations of disorder we set the Hamiltonian parameters
so that the (final) set $\{\alpha_j\}$ displays a patchy pattern, where
blotches with $\alpha_j \ne 0$ are separated by regions where
$\alpha_j$'s virtually vanish. This has the effect of quenching the
SF fraction \cite{A:disMF2d,A:bmpvN}. Conversely, the condensate fraction
is small yet significant, being basically 
$f_{\rm C}= {\cal N}^{-1}\,\sum_j|\alpha_j|^2$. 
The resulting system is not SF,
yet it retains some degree of coherence. Since this phase arises in the presence of disorder and is neither SF nor MI, we identify it with the 
BG \cite{A:disMF2d}. This is illustrated in Figs. \ref{F:fig1} and 
\ref{F:fig3}, referring to the uniform and binary distribution of
the disordered local potentials, respectively. 

Technically, the system is initialized by choosing a set $\{\alpha_j\}$
which clearly does not satisfy the above described constraint.
Subsequently, the sites of the system are addressed singularly and the
relevant local order parameter is changed in order to minimize the energy
while meeting the constraint. At each iteration all of the sites undergo
one of these local moves.
%
%
During this process the chemical potential
is adjusted in order to achieve the desired boson population. We assume that the system has reached convergence when 
$max_j \left(\left|\alpha^{s+1}_j-\alpha^{s}_j
\right|\right)\leq 2.5\cdot 10^{-4}$,
%
%
where $s$ and $s+1$ denote subsequent steps in the iteration process.
When this condition is achieved, the constraint on the total population
is met with a much higher precision.

In principle the local order parameters are complex numbers,
but we clearly observe that the state minimizing the system
energy is real. 
%
%
More precisely, the lowest energy is attained by many configurations characterized by the same local moduli and a different global phase,
$\alpha_j = |\alpha_j| e^{i \varphi}$. Much interestingly, we also
find a large number of very low lying states characterized by
basically the same patchy $|\alpha_j|$ distribution as the
ground-state(s) and a non-trivial phase distribution. The phase of
the local parameters is not constant throughout the lattice, but only
inside each non-vanishing $|\alpha_j|$ patch. Different patches feature 
in general a different ``global'' phase.

\begin{figure}[t!]
\begin{center}
\includegraphics[width= 18cm]{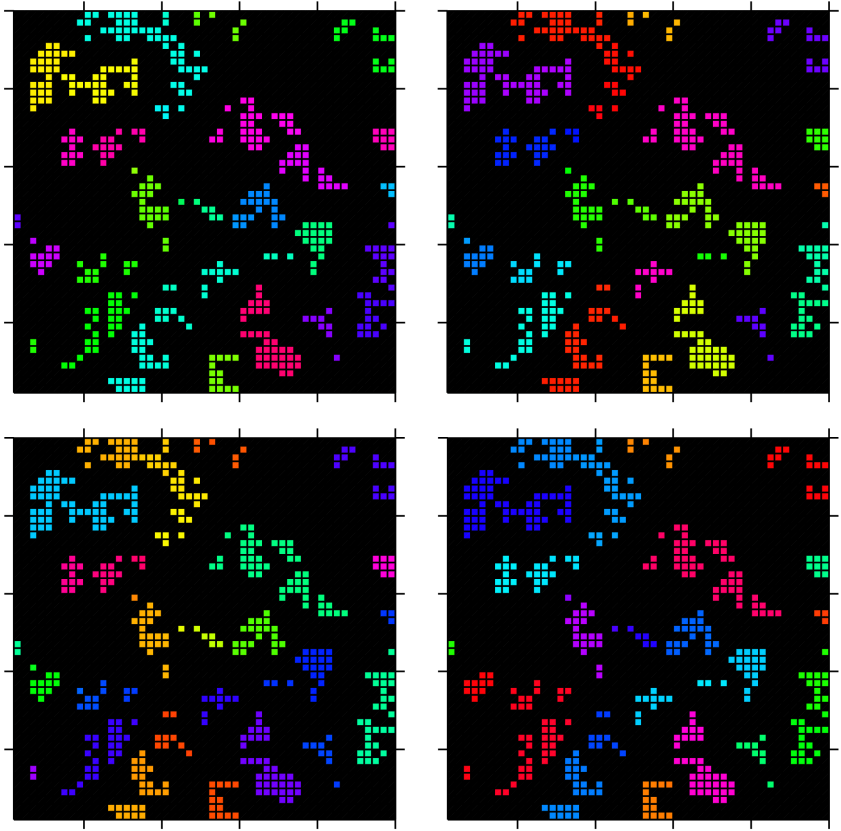}
\caption{\label{F:fig2}
Local phase patterns for the same situation discussed in Fig. \ref{F:fig1}. The four configurations have been obtained starting from different initial random phase patterns. The phases range from $0$ to $2 \pi$ as described by the lowermost colorbar in Fig. \ref{F:fig1}. Black areas correspond to region of vanishing $|\langle a_j \rangle|^2$. As discussed in the text, the $\langle n_j \rangle$ and $|\langle a_j \rangle|^2$ distributions of these complex configurations are virtually undistinguishable from those of the real configuration, displayed in the rightmost panels of Fig. \ref{F:fig1}.}
\end{center}
\end{figure}

\begin{figure}[t!]
\begin{center}
\includegraphics[width= 18cm]{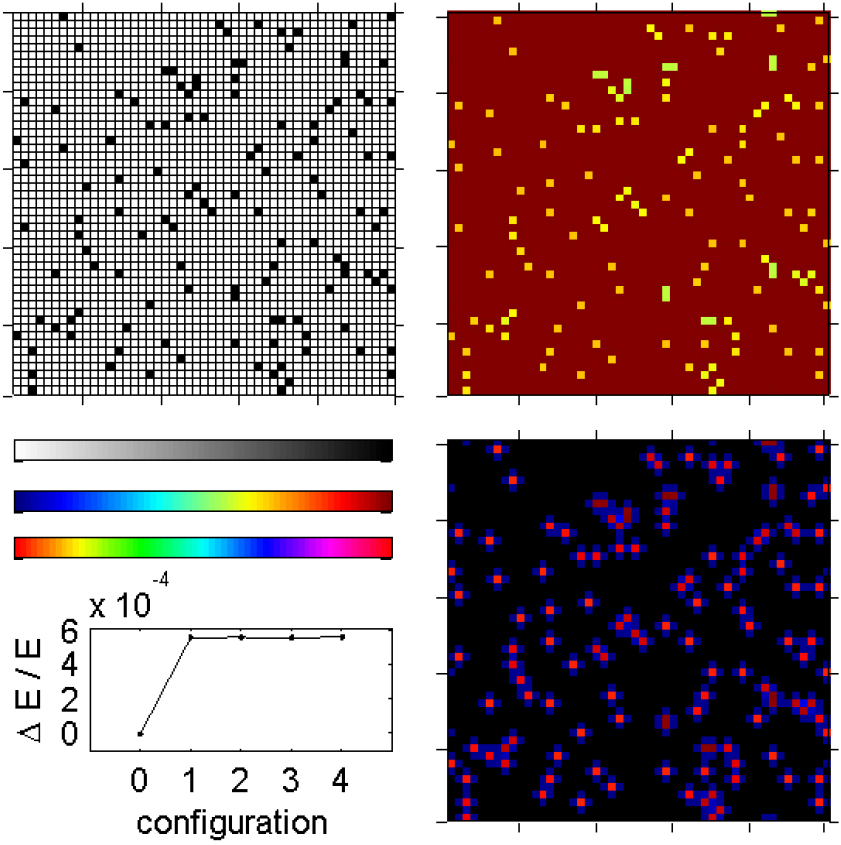}
\caption{\label{F:fig3}
Binary  random potential ($\Delta = 0.5$, $128$ impurities) on a $50 \times 50$ lattice with periodic boundary conditions containing $N=2455$ bosons with hopping amplitude $J=1.8\cdot 10^{-2}$. The panels have the same meaning as in Fig. \ref{F:fig1}. The color ranges for the local potentials (top colorbar), densities and order parameters (middle colorbar) are $[0,0.5]$, $[0,1]$,$[0,0.263]$, respectively.}
\end{center}
\end{figure}

Let us now discuss our results in more detail. As we say, Fig. \ref{F:fig1} addresses the case of uniformly distributed local random potentials. 
The potential landscape $v_j \in [-0.2 U, 0.2 U]$ is shown in the 
upper left panel of the figure. The corresponding colorbar is the 
topmost one in the lower left quadrant of the figure. We choose to 
load the $50\times50$ lattice with ${\cal N}=2650$ bosons, corresponding
to a filling slightly larger than unity. 
We find that at $J/U = 6\cdot10^{-3}$ the ground state is such that
most of the lattice sites have a local population of one single boson, 
with a very small deviation. This is clear from the topmost left panel, 
showing the local population, where the colormap ranges from $0$ to $2$ 
and unitary population corresponds to a green hue.
The local order parameter $\alpha_j$ virtually vanishes at these ``highly squeezed'' sites, as it is illustrated in the lower right panel of Fig. \ref{F:fig1}. The corresponding colormap is the same as above, but the color range is $\alpha_j \in [0,0.527]$. The sites with virtually vanishing local order parameters have been assigned a black hue in order to highlight the blotchy pattern of the nonzero $\alpha_j$'s. 
 The patches of nonzero $\alpha_j$'s add up to a small yet defininitely nonvanishing condensate fraction, whereas the ``sea'' of vanishing local order parameter results in a virtually zero $f_{\rm S}$. As a side effect, this
 ``sea'' allows for the existence of very low-energy complex-$\alpha_j$ configurations characterized by the same local population and $|\alpha_j|$ pattern as the ground-state, but different phase pattern. Figure \ref{F:fig2} shows such phase patterns for four of these low-lying configurations. The color key for these figures is encoded in the lowermost colorbar in the lower left quadrant of Fig. \ref{F:fig1}, which ranges from $0$ to $2\pi$. The plot in the same quadrant shows the relative energy difference between the ground-state (0) and the four configurations in Fig. \ref{F:fig2} (1-4). Note that the islands of nonzero local order parameters in Fig. \ref{F:fig2} have the same boundaries, and that all the sites belonging to the same island have virtually the same phase $-i\log \left(\alpha_j/|\alpha_j|\right)$. Different islands in the same configuration or even the same island in different configurations have in general different ``island-phases'', recognizable as different colors.

These low-lying states appear to be quite robust to our algorithm searching for the ground-state of the system, and in this sense we refer to them as {\it metastable states}. Our algorithm gets stuck at these low-energy configurations virtually indefinitely
and, in this sense, we refer to them as metastable states. A similar behaviour has been recently observed in related systems, even in the absence of disorder \cite{A:Menot,Roscilde_PRL_98_190402}.

\begin{figure}[t!]
\begin{center}
\includegraphics[width= 18cm]{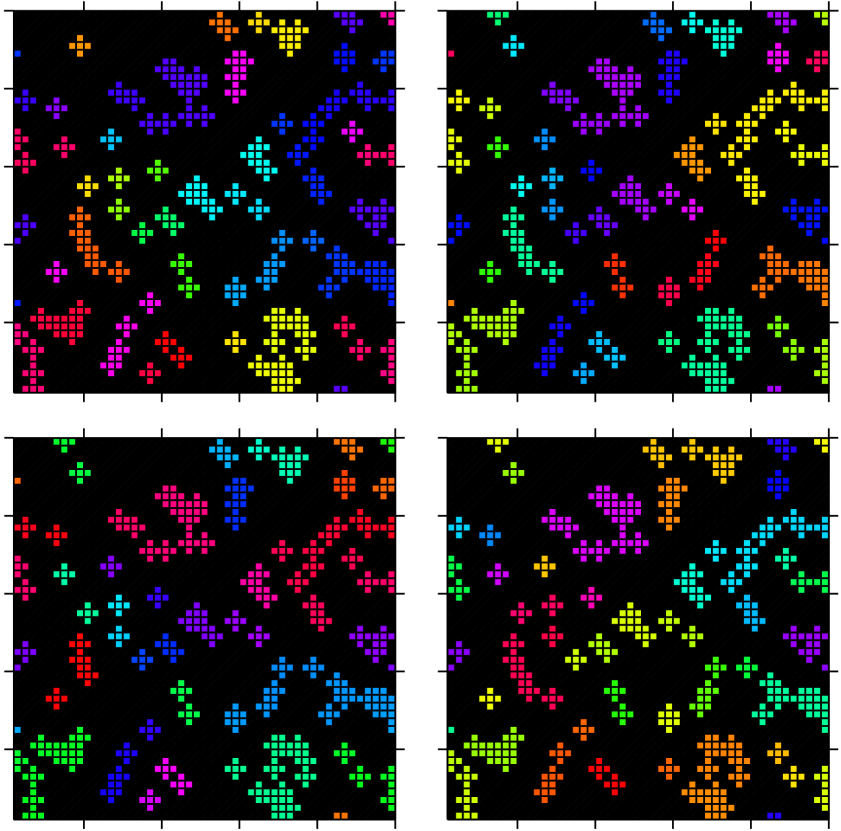}
\caption{\label{F:fig4} 
Local phase patterns for the same situation discussed in Fig. \ref{F:fig3}. The four configurations have been obtained starting from different initial random phase patterns. The phases range from $0$ to $2 \pi$ as described by the lowermost colorbar in Fig. \ref{F:fig3}. Black areas correspond to region of vanishing $|\langle a_j \rangle|^2$. The situation is analogous to that of the uniformly distributed random local potentials discussed in fig \ref{F:fig2}.
}
\end{center}
\end{figure}

Figure \ref{F:fig3} shows a similar situation on a $50\times50$ lattice where a binary disorder is present. In 128 out of the 2500 total sites the local potential equals $0.5 U$, whereas it is zero in the remaining sites of the lattice. The local potential landscape is illustrated in the top left quadrant of Fig. \ref{F:fig3}. It could describe the effect of 128 atoms of a second atomic species, whose position is ``frozen'' due to a greatly quenched hopping amplitude, which interact repulsively with the atoms of the bosonic species \cite{A:Ospelkaus,A:Gavish}.
For the latter we once again choose a filling close to unity, yet slightly lower, ${\cal N} = 2455$, and $J/U = 1.8\cdot 10^{-2}$.
Once again the local population of the ground state is $1$ with a very small variance at most of the lattice sites. The remaining sites, which are those where the impurities are localized, and possibly their nearest neighbours, have a smaller population and a larger population variance. This is clear from the upper right panel in Fig. \ref{F:fig3} (the relevant colormap is the same as in Fig. \ref{F:fig1}, with the only difference that it ranges in $[0, 1]$ so that in this case unitary filling corresponds to a dark red hue). The lower left panel of Fig. \ref{F:fig3} shows that nonzero local order parameters appear at the impurity sites and possibly at their nearest neighbours. The remaining sites of the lattice have a virtually zero local order parameters, which produces the same effect as discussed above. Once again we find low-lying states characterized by virtyally the same blotchy distribution of the (modulus of the) local order pameter, but different phases on different nonvanishing-$\alpha_j$ islands. This is clearly illustrated in Fig. \ref{F:fig4}, whose color key is the same as Fig. \ref{F:fig2}.

As  $J/U$ is increased the nonzero-local-order-parameter islands swell, and eventually melt together. This allows a superfluid current to flow through the system, and the ground-state is not any more in a BG phase, but rather in a SF phase. Under these circumstances we find no low-lying states characterized by a random phase pattern of the local order parameter. This either means that such states do not exist or that they are not robust to our algorithm as those observed in the BG phase. Even starting our algorithm with local order parameters exhibiting a random phase pattern ends up in a ``real'' ground-state (modulo a global U$(1)$ symmetry). When convergence is achieved the phases of the local order parameters are the same to a good approximation.

\section{Discussion}
We consider a two-dimensionl BH model with random local potentials, focusing on two different disorder distributions, uniform and binary.
We show that in the BG phase  many low-energy metastable configurations exist with a slightly larger energy than the ground-state. These configurations are characterized by the same blotchy pattern of local order parameters as the ground state. However, while in the latter the local order parameters have the same global phase, in the latter each nonvanishing-$\alpha_j$ island is characterized by a different constant phase. This situation is strongly reminiscent of the energy landscape of glassy systems. 

We observe that the random phase pattern characterizing the low-lying states in the BG phase could account for the experimental observation that the BG appears to be lacking coherence \cite{A:Fall07}. Indeed this is apparently in contrast with our characterization of the BG as a non-superfluid yet coherent phase.
Clearly, one should take into account the fact that the spatial arrangement of the coherent blotches characterizing the BG is very complex, following from the random nature of the local potential pattern. The interference pattern produced by a large number of coherent yet randomly localized sources is very likely not so different from that of a set of incoherent sources.
One could in principle think of ``factoring out'' the geometric contribution to the interference pattern. For instance, in the binary disorder case one could observe that the location of the coherent sources is basically the same as that of the random impurities. If the location of the latter is somehow known, it should be possible to recognize whether an interference pattern is actually incoherent or it looks so due to the random location of the coherent sources.

However the incoherence of the interference pattern could arise from the ``glassiness'' of the BG. Due e. g. to imperfect cooling the system could be on a slightly excited state instead of the ground-state. If this is the case, correlating the interference pattern with the distribution of coherent source could prove useless, because anyway each of these sources could be characterized by a different unknown phase.

\vfill\eject

\begin{thebibliography}{32}
\expandafter\ifx\csname natexlab\endcsname\relax\def\natexlab#1{#1}\fi
\expandafter\ifx\csname bibnamefont\endcsname\relax
  \def\bibnamefont#1{#1}\fi
\expandafter\ifx\csname bibfnamefont\endcsname\relax
  \def\bibfnamefont#1{#1}\fi
\expandafter\ifx\csname citenamefont\endcsname\relax
  \def\citenamefont#1{#1}\fi
\expandafter\ifx\csname url\endcsname\relax
  \def\url#1{\texttt{#1}}\fi
\expandafter\ifx\csname urlprefix\endcsname\relax\def\urlprefix{URL }\fi
\providecommand{\bibinfo}[2]{#2}
\providecommand{\eprint}[2][]{\url{#2}}

\bibitem[{\citenamefont{Fisher \emph{et~al.}}(1989)\citenamefont{Fisher,
  Weichman, Grinstein, and Fisher}}]{A:Fisher}

\bibinfo{author}{\bibfnamefont{M.~P.~A.} \bibnamefont{Fisher}},
  \bibinfo{author}{\bibfnamefont{P.~B.} \bibnamefont{Weichman}},
  \bibinfo{author}{\bibfnamefont{G.}~\bibnamefont{Grinstein}},
  \bibnamefont{and} \bibinfo{author}{\bibfnamefont{D.~S.}
  \bibnamefont{Fisher}}, \bibinfo{journal}{Phys. Rev. B}
  \textbf{\bibinfo{volume}{40}}, \bibinfo{pages}{546} (\bibinfo{year}{1989}).

\bibitem[{\citenamefont{Scalettar \emph{et~al.}}(1991)\citenamefont{Scalettar,
  Batrouni, and Zimanyi}}]{A:Scalettar91}

\bibinfo{author}{\bibfnamefont{R.~T.} \bibnamefont{Scalettar}},
  \bibinfo{author}{\bibfnamefont{G.~G.} \bibnamefont{Batrouni}},
  \bibnamefont{and} \bibinfo{author}{\bibfnamefont{G.~T.}
  \bibnamefont{Zimanyi}}, \bibinfo{journal}{Phys. Rev. Lett.}
  \textbf{\bibinfo{volume}{66}}, \bibinfo{pages}{3144} (\bibinfo{year}{1991}).

\bibitem[{\citenamefont{Freericks and Monien}(1996)}]{A:Freericks96}

\bibinfo{author}{\bibfnamefont{J.~K.} \bibnamefont{Freericks}}
  \bibnamefont{and} \bibinfo{author}{\bibfnamefont{H.}~\bibnamefont{Monien}},
  \bibinfo{journal}{Phys. Rev. B} \textbf{\bibinfo{volume}{53}},
  \bibinfo{pages}{2691} (\bibinfo{year}{1996}).

\bibitem[{\citenamefont{Rapsch \emph{et~al.}}(1999)\citenamefont{Rapsch,
  Schollwšock, and Zwerger}}]{A:Rapsch99}

\bibinfo{author}{\bibfnamefont{S.}~\bibnamefont{Rapsch}},
  \bibinfo{author}{\bibfnamefont{U.}~\bibnamefont{Schollwšock}},
  \bibnamefont{and} \bibinfo{author}{\bibfnamefont{W.}~\bibnamefont{Zwerger}},
  \bibinfo{journal}{Europhys. Lett.} \textbf{\bibinfo{volume}{46}},
  \bibinfo{pages}{559} (\bibinfo{year}{1999}).

\bibitem[{\citenamefont{Gimperlein
  \emph{et~al.}}(2005)\citenamefont{Gimperlein, Wessel, Schmiedmayer, and
  Santos}}]{A:Gimper05}

\bibinfo{author}{\bibfnamefont{H.}~\bibnamefont{Gimperlein}},
  \bibinfo{author}{\bibfnamefont{S.}~\bibnamefont{Wessel}},
  \bibinfo{author}{\bibfnamefont{J.}~\bibnamefont{Schmiedmayer}},
  \bibnamefont{and} \bibinfo{author}{\bibfnamefont{L.}~\bibnamefont{Santos}},
  \bibinfo{journal}{Phys. Rev. Lett.} \textbf{\bibinfo{volume}{95}},
  \bibinfo{pages}{170401} (\bibinfo{year}{2005}).

\bibitem[{\citenamefont{Martino \emph{et~al.}}(2005)\citenamefont{Martino,
  Thorwart, Egger, and Graham}}]{A:DeMart05}

\bibinfo{author}{\bibfnamefont{A.~D.} \bibnamefont{Martino}},
  \bibinfo{author}{\bibfnamefont{M.}~\bibnamefont{Thorwart}},
  \bibinfo{author}{\bibfnamefont{R.}~\bibnamefont{Egger}}, \bibnamefont{and}
  \bibinfo{author}{\bibfnamefont{R.}~\bibnamefont{Graham}},
  \bibinfo{journal}{Phys. Rev. Lett.} \textbf{\bibinfo{volume}{94}},
  \bibinfo{pages}{060402} (\bibinfo{year}{2005}).

\bibitem[{\citenamefont{Yukalov and Graham}(2007)}]{A:Yuka07}

\bibinfo{author}{\bibfnamefont{V.}~\bibnamefont{Yukalov}} \bibnamefont{and}
  \bibinfo{author}{\bibfnamefont{R.}~\bibnamefont{Graham}},
  \bibinfo{journal}{Phys. Rev. A} \textbf{\bibinfo{volume}{75}},
  \bibinfo{pages}{023619} (\bibinfo{year}{2007}).

\bibitem[{\citenamefont{Mering and Fleischhauer}(2007)}]{A:Mering07}

\bibinfo{author}{\bibfnamefont{A.}~\bibnamefont{Mering}} \bibnamefont{and}
  \bibinfo{author}{\bibfnamefont{M.}~\bibnamefont{Fleischhauer}},
  \bibinfo{journal}{arXiv:0709.2386v2}  (\bibinfo{year}{2007}).

\bibitem[{\citenamefont{Krauth \emph{et~al.}}(1992)\citenamefont{Krauth,
  Caffarel, and Bouchaud}}]{A:Krauth92}

\bibinfo{author}{\bibfnamefont{W.}~\bibnamefont{Krauth}},
  \bibinfo{author}{\bibfnamefont{M.}~\bibnamefont{Caffarel}}, \bibnamefont{and}
  \bibinfo{author}{\bibfnamefont{J.-P.} \bibnamefont{Bouchaud}},
  \bibinfo{journal}{Phys. Rev. B} \textbf{\bibinfo{volume}{45}},
  \bibinfo{pages}{3137} (\bibinfo{year}{1992}).

\bibitem[{\citenamefont{Singh and Rokhsar}(1994)}]{A:Singh94}

\bibinfo{author}{\bibfnamefont{K.~G.} \bibnamefont{Singh}} \bibnamefont{and}
  \bibinfo{author}{\bibfnamefont{D.~S.} \bibnamefont{Rokhsar}},
  \bibinfo{journal}{Phys. Rev. B} \textbf{\bibinfo{volume}{49}},
  \bibinfo{pages}{9013} (\bibinfo{year}{1994}).

\bibitem[{\citenamefont{Sengupta and Haas}(2007)}]{A:Sengupta99}

\bibinfo{author}{\bibfnamefont{P.}~\bibnamefont{Sengupta}} \bibnamefont{and}
  \bibinfo{author}{\bibfnamefont{S.}~\bibnamefont{Haas}},
  \bibinfo{journal}{Phys. Rev. Lett.} \textbf{\bibinfo{volume}{99}},
  \bibinfo{pages}{050403} (\bibinfo{year}{2007}).

\bibitem[{\citenamefont{Balabanyan
  \emph{et~al.}}(2005)\citenamefont{Balabanyan, Prokof'ev, and
  Svistunov}}]{A:Balab05}

\bibinfo{author}{\bibfnamefont{K.~G.} \bibnamefont{Balabanyan}},
  \bibinfo{author}{\bibfnamefont{N.}~\bibnamefont{Prokof'ev}},
  \bibnamefont{and}
  \bibinfo{author}{\bibfnamefont{B.}~\bibnamefont{Svistunov}},
  \bibinfo{journal}{Phys. Rev. Lett.} \textbf{\bibinfo{volume}{95}},
  \bibinfo{pages}{055701} (\bibinfo{year}{2005}).

\bibitem[{\citenamefont{Bernardet \emph{et~al.}}(2002)\citenamefont{Bernardet,
  Batrouni, and Troyer}}]{A:Bernad02}

\bibinfo{author}{\bibfnamefont{K.}~\bibnamefont{Bernardet}},
  \bibinfo{author}{\bibfnamefont{G.~G.} \bibnamefont{Batrouni}},
  \bibnamefont{and} \bibinfo{author}{\bibfnamefont{M.}~\bibnamefont{Troyer}},
  \bibinfo{journal}{Phys. Rev. B} \textbf{\bibinfo{volume}{66}},
  \bibinfo{pages}{054520} (\bibinfo{year}{2002}).

\bibitem[{\citenamefont{Prokof'ev and Svistunov}(2004)}]{A:Prokov04}

\bibinfo{author}{\bibfnamefont{N.}~\bibnamefont{Prokof'ev}} \bibnamefont{and}
  \bibinfo{author}{\bibfnamefont{B.}~\bibnamefont{Svistunov}},
  \bibinfo{journal}{Phys. Rev. Lett.} \textbf{\bibinfo{volume}{92}},
  \bibinfo{pages}{015703} (\bibinfo{year}{2004}).

\bibitem[{\citenamefont{Lye \emph{et~al.}}(2005)\citenamefont{Lye, Fallani,
  Modugno, Wiersma, Fort, and Inguscio}}]{A:Lye}

\bibinfo{author}{\bibfnamefont{J.~E.} \bibnamefont{Lye}},
  \bibinfo{author}{\bibfnamefont{L.}~\bibnamefont{Fallani}},
  \bibinfo{author}{\bibfnamefont{M.}~\bibnamefont{Modugno}},
  \bibinfo{author}{\bibfnamefont{D.~S.} \bibnamefont{Wiersma}},
  \bibinfo{author}{\bibfnamefont{C.}~\bibnamefont{Fort}}, \bibnamefont{and}
  \bibinfo{author}{\bibfnamefont{M.}~\bibnamefont{Inguscio}},
  \bibinfo{journal}{Phys. Rev. Lett.} \textbf{\bibinfo{volume}{95}},
  \bibinfo{pages}{070401} (\bibinfo{year}{2005}).

\bibitem[{\citenamefont{Schulte \emph{et~al.}}(2005)\citenamefont{Schulte,
  Drenkelforth, Kruse, Ertmer, Arlt, Sacha, Zakrzewski, and
  Lewenstein}}]{A:Schulte}

\bibinfo{author}{\bibfnamefont{T.}~\bibnamefont{Schulte}},
  \bibinfo{author}{\bibfnamefont{S.}~\bibnamefont{Drenkelforth}},
  \bibinfo{author}{\bibfnamefont{J.}~\bibnamefont{Kruse}},
  \bibinfo{author}{\bibfnamefont{W.}~\bibnamefont{Ertmer}},
  \bibinfo{author}{\bibfnamefont{J.}~\bibnamefont{Arlt}},
  \bibinfo{author}{\bibfnamefont{K.}~\bibnamefont{Sacha}},
  \bibinfo{author}{\bibfnamefont{J.}~\bibnamefont{Zakrzewski}},
  \bibnamefont{and}
  \bibinfo{author}{\bibfnamefont{M.}~\bibnamefont{Lewenstein}},
  \bibinfo{journal}{Phys. Rev. Lett.} \textbf{\bibinfo{volume}{95}},
  \bibinfo{pages}{170411} (\bibinfo{year}{2005}).

\bibitem[{\citenamefont{Damski \emph{et~al.}}(2003)\citenamefont{Damski,
  Zakrzewski, Santos, Zoller, and Lewenstein}}]{A:Damski2003}

\bibinfo{author}{\bibfnamefont{B.}~\bibnamefont{Damski}},
  \bibinfo{author}{\bibfnamefont{J.}~\bibnamefont{Zakrzewski}},
  \bibinfo{author}{\bibfnamefont{L.}~\bibnamefont{Santos}},
  \bibinfo{author}{\bibfnamefont{P.}~\bibnamefont{Zoller}}, \bibnamefont{and}
  \bibinfo{author}{\bibfnamefont{M.}~\bibnamefont{Lewenstein}},
  \bibinfo{journal}{Phys. Rev. Lett.} \textbf{\bibinfo{volume}{91}},
  \bibinfo{pages}{080403} (\bibinfo{year}{2003}).

\bibitem[{\citenamefont{Fallani \emph{et~al.}}(2007)\citenamefont{Fallani, Lye,
  Guarrera, Fort, and Inguscio}}]{A:Fall07}

\bibinfo{author}{\bibfnamefont{L.}~\bibnamefont{Fallani}},
  \bibinfo{author}{\bibfnamefont{J.~E.} \bibnamefont{Lye}},
  \bibinfo{author}{\bibfnamefont{V.}~\bibnamefont{Guarrera}},
  \bibinfo{author}{\bibfnamefont{C.}~\bibnamefont{Fort}}, \bibnamefont{and}
  \bibinfo{author}{\bibfnamefont{M.}~\bibnamefont{Inguscio}},
  \bibinfo{journal}{Phys. Rev. Lett.} \textbf{\bibinfo{volume}{98}},
  \bibinfo{pages}{130404} (\bibinfo{year}{2007}).

\bibitem[{\citenamefont{Gavish and Castin}(2005)}]{A:Gavish}

\bibinfo{author}{\bibfnamefont{U.}~\bibnamefont{Gavish}} \bibnamefont{and}
  \bibinfo{author}{\bibfnamefont{Y.}~\bibnamefont{Castin}},
  \bibinfo{journal}{Phys. Rev. Lett.} \textbf{\bibinfo{volume}{95}},
  \bibinfo{eid}{020401} (pages~\bibinfo{numpages}{4}) (\bibinfo{year}{2005}).

\bibitem[{\citenamefont{Ospelkaus \emph{et~al.}}(2006)\citenamefont{Ospelkaus,
  Ospelkaus, Wille, Succo, Ernst, Sengstock, and Bongs}}]{A:Ospelkaus}

\bibinfo{author}{\bibfnamefont{S.}~\bibnamefont{Ospelkaus}},
  \bibinfo{author}{\bibfnamefont{C.}~\bibnamefont{Ospelkaus}},
  \bibinfo{author}{\bibfnamefont{O.}~\bibnamefont{Wille}},
  \bibinfo{author}{\bibfnamefont{M.}~\bibnamefont{Succo}},
  \bibinfo{author}{\bibfnamefont{P.}~\bibnamefont{Ernst}},
  \bibinfo{author}{\bibfnamefont{K.}~\bibnamefont{Sengstock}},
  \bibnamefont{and} \bibinfo{author}{\bibfnamefont{K.}~\bibnamefont{Bongs}},
  \bibinfo{journal}{Phys. Rev. Lett.} \textbf{\bibinfo{volume}{96}},
  \bibinfo{pages}{180403} (\bibinfo{year}{2006}).

\bibitem[{\citenamefont{Buonsante
  \emph{et~al.}}(2007{\natexlab{a}})\citenamefont{Buonsante, Massel, Penna, and
  Vezzani}}]{A:RandhopMF}

\bibinfo{author}{\bibfnamefont{P.}~\bibnamefont{Buonsante}},
  \bibinfo{author}{\bibfnamefont{F.}~\bibnamefont{Massel}},
  \bibinfo{author}{\bibfnamefont{V.}~\bibnamefont{Penna}}, \bibnamefont{and}
  \bibinfo{author}{\bibfnamefont{A.}~\bibnamefont{Vezzani}},
  \bibinfo{journal}{Laser Phys.} \textbf{\bibinfo{volume}{17}},
  \bibinfo{pages}{538} (\bibinfo{year}{2007}{\natexlab{a}}).

\bibitem[{\citenamefont{Sheshadri \emph{et~al.}}(1995)\citenamefont{Sheshadri,
  Krishnamurthy, Pandit, and Ramakrishnan}}]{A:Sheshadri1995}

\bibinfo{author}{\bibfnamefont{K.}~\bibnamefont{Sheshadri}},
  \bibinfo{author}{\bibfnamefont{H.~R.} \bibnamefont{Krishnamurthy}},
  \bibinfo{author}{\bibfnamefont{R.}~\bibnamefont{Pandit}}, \bibnamefont{and}
  \bibinfo{author}{\bibfnamefont{T.~V.} \bibnamefont{Ramakrishnan}},
  \bibinfo{journal}{Phys. Rev. Lett.} \textbf{\bibinfo{volume}{75}},
  \bibinfo{pages}{4075} (\bibinfo{year}{1995}).

\bibitem[{\citenamefont{Buonsante
  \emph{et~al.}}(2007{\natexlab{b}})\citenamefont{Buonsante, Penna, Vezzani,
  and Blakie}}]{A:disMF1d}

\bibinfo{author}{\bibfnamefont{P.}~\bibnamefont{Buonsante}},
  \bibinfo{author}{\bibfnamefont{V.}~\bibnamefont{Penna}},
  \bibinfo{author}{\bibfnamefont{A.}~\bibnamefont{Vezzani}}, \bibnamefont{and}
  \bibinfo{author}{\bibfnamefont{P.~B.} \bibnamefont{Blakie}},
  \bibinfo{journal}{Phys. Rev. A} \textbf{\bibinfo{volume}{76}},
  \bibinfo{pages}{011602(R)} (\bibinfo{year}{2007}{\natexlab{b}}).

\bibitem[{\citenamefont{Buonsante
  \emph{et~al.}}(2007{\natexlab{c}})\citenamefont{Buonsante, Massel, Penna, and
  Vezzani}}]{A:disMF2d}

\bibinfo{author}{\bibfnamefont{P.}~\bibnamefont{Buonsante}},
  \bibinfo{author}{\bibfnamefont{F.}~\bibnamefont{Massel}},
  \bibinfo{author}{\bibfnamefont{V.}~\bibnamefont{Penna}}, \bibnamefont{and}
  \bibinfo{author}{\bibfnamefont{A.}~\bibnamefont{Vezzani}},
  \bibinfo{journal}{J. Phys. B-FTC} \textbf{\bibinfo{volume}{76}},
  \bibinfo{pages}{F265} (\bibinfo{year}{2007}{\natexlab{c}}).

\bibitem[{\citenamefont{Krutitsky \emph{et~al.}}(2006)\citenamefont{Krutitsky,
  A., and R.}}]{A:Krutitsky}

\bibinfo{author}{\bibfnamefont{K.~V.} \bibnamefont{Krutitsky}},
  \bibinfo{author}{\bibfnamefont{P.}~\bibnamefont{A.}}, \bibnamefont{and}
  \bibinfo{author}{\bibfnamefont{G.}~\bibnamefont{R.}}, \bibinfo{journal}{New
  J. Phys.} \textbf{\bibinfo{volume}{8}}, \bibinfo{pages}{187}
  (\bibinfo{year}{2006}).

\bibitem[{\citenamefont{Clement \emph{et~al.}}(2005)\citenamefont{Clement,
  Varon, Hugbart, Retter, Bouyer, Sanchez-Palencia, Gangardt, Shlyapnikov, and
  Aspect}}]{A:Clement05}

\bibinfo{author}{\bibfnamefont{D.}~\bibnamefont{Clement}},
  \bibinfo{author}{\bibfnamefont{A.~F.} \bibnamefont{Varon}},
  \bibinfo{author}{\bibfnamefont{M.}~\bibnamefont{Hugbart}},
  \bibinfo{author}{\bibfnamefont{J.~A.} \bibnamefont{Retter}},
  \bibinfo{author}{\bibfnamefont{P.}~\bibnamefont{Bouyer}},
  \bibinfo{author}{\bibfnamefont{L.}~\bibnamefont{Sanchez-Palencia}},
  \bibinfo{author}{\bibfnamefont{D.~M.} \bibnamefont{Gangardt}},
  \bibinfo{author}{\bibfnamefont{G.~V.} \bibnamefont{Shlyapnikov}},
  \bibnamefont{and} \bibinfo{author}{\bibfnamefont{A.}~\bibnamefont{Aspect}},
  \bibinfo{journal}{Phys. Rev. Lett.} \textbf{\bibinfo{volume}{95}},
  \bibinfo{pages}{170409} (\bibinfo{year}{2005}).

\bibitem[{\citenamefont{Roth and Burnett}(2003)}]{A:Roth}

\bibinfo{author}{\bibfnamefont{R.}~\bibnamefont{Roth}} \bibnamefont{and}
  \bibinfo{author}{\bibfnamefont{K.}~\bibnamefont{Burnett}},
  \bibinfo{journal}{Phys. Rev. A} \textbf{\bibinfo{volume}{68}},
  \bibinfo{pages}{023604} (\bibinfo{year}{2003}).

\bibitem[{\citenamefont{Jaksch \emph{et~al.}}(2002)\citenamefont{Jaksch,
  Venturi, Cirac, Williams, and Zoller}}]{A:Jaksch2}

\bibinfo{author}{\bibfnamefont{D.}~\bibnamefont{Jaksch}},
  \bibinfo{author}{\bibfnamefont{V.}~\bibnamefont{Venturi}},
  \bibinfo{author}{\bibfnamefont{J.~I.} \bibnamefont{Cirac}},
  \bibinfo{author}{\bibfnamefont{C.~J.} \bibnamefont{Williams}},
  \bibnamefont{and} \bibinfo{author}{\bibfnamefont{P.}~\bibnamefont{Zoller}},
  \bibinfo{journal}{Phys. Rev. Lett.} \textbf{\bibinfo{volume}{89}},
  \bibinfo{pages}{040402} (\bibinfo{year}{2002}).

\bibitem[{\citenamefont{Snoek and Hofstetter}(2007)}]{A:Snoek07}

\bibinfo{author}{\bibfnamefont{M.}~\bibnamefont{Snoek}} \bibnamefont{and}
  \bibinfo{author}{\bibfnamefont{W.}~\bibnamefont{Hofstetter}},
  \bibinfo{journal}{Phys. Rev. A} \textbf{\bibinfo{volume}{76}},
  \bibinfo{pages}{051603(R)} (\bibinfo{year}{2007}).

\bibitem[{\citenamefont{Buonsante
  \emph{et~al.}}(2007{\natexlab{d}})\citenamefont{Buonsante, Massel, Penna, and
  Vezzani}}]{A:bmpvN}

\bibinfo{author}{\bibfnamefont{P.}~\bibnamefont{Buonsante}},
  \bibinfo{author}{\bibfnamefont{F.}~\bibnamefont{Massel}},
  \bibinfo{author}{\bibfnamefont{V.}~\bibnamefont{Penna}}, \bibnamefont{and}
  \bibinfo{author}{\bibfnamefont{A.}~\bibnamefont{Vezzani}},
  \bibinfo{journal}{in preparation}  (\bibinfo{year}{2007}{\natexlab{d}}).

\bibitem[{\citenamefont{Menotti \emph{et~al.}}(2007)\citenamefont{Menotti,
  Trefzger, and Lewenstein}}]{A:Menot}

\bibinfo{author}{\bibfnamefont{C.}~\bibnamefont{Menotti}},
  \bibinfo{author}{\bibfnamefont{C.}~\bibnamefont{Trefzger}}, \bibnamefont{and}
  \bibinfo{author}{\bibfnamefont{M.}~\bibnamefont{Lewenstein}},
  \bibinfo{journal}{Phys. Rev. Lett.} \textbf{\bibinfo{volume}{98}},
  \bibinfo{pages}{235301} (\bibinfo{year}{2007}).

\bibitem[{\citenamefont{Roscilde and Cirac}(2007)}]{Roscilde_PRL_98_190402}

\bibinfo{author}{\bibfnamefont{T.}~\bibnamefont{Roscilde}} \bibnamefont{and}
  \bibinfo{author}{\bibfnamefont{J.~I.} \bibnamefont{Cirac}},
  \bibinfo{journal}{Phys. Rev. Lett.} \textbf{\bibinfo{volume}{98}},
  \bibinfo{pages}{190402} (\bibinfo{year}{2007}).

\end{thebibliography}
\end{document}